\documentclass[a4paper,12pt]{article}
\usepackage{epsfig}
\usepackage{subfig}
\usepackage{amssymb}
\begin{document}
\thispagestyle{empty}

\newcommand{\etal}  {{\it{et al.}}}  
\def\Journal#1#2#3#4{{#1} {\bf #2}, #3 (#4)}
\def\PRD{Phys.\ Rev.\ D}
\def\NIMA{Nucl.\ Instrum.\ Methods A}
\def\PRL{Phys.\ Rev.\ Lett.\ }
\def\PLB{Phys.\ Lett.\ B}
\def\EPJ{Eur.\ Phys.\ J}
\def\IEEETNS{IEEE Trans.\ Nucl.\ Sci.\ }
\def\CPCD{Comput.\ Phys.\ Commun.\ }

\smallskip

\bigskip
\bigskip

{\Large\bf
\begin{center}
On decays of $Z^{\prime}$ into unparticle stuff
\end{center}
}
\vspace{1 cm}

\begin{center}
{\large G.A. Kozlov  }
\end{center}
\begin{center}
\noindent
 { Bogolyubov Laboratory of Theoretical Physics\\
 Joint Institute for Nuclear Research,\\
 Joliot Curie st., 6, Dubna, Moscow region, 141980 Russia  }
\end{center}
\begin{center}
{\large I.N. Gorbunov}
\end{center}
\begin{center}
\noindent
 {University Centre\\
 Joint Institute for Nuclear Research,\\
 Joliot Curie st., 6, Dubna, Moscow region, 141980 Russia  }
\end{center}
\vspace{0.5 cm}

 \begin{abstract}
 \noindent
 {We study the decay of a $Z^{\prime}$ - boson into  $U$ -unparticle  and a  photon. The extended Landau-Yang theorem is used. The clear photon signal would make the decay $Z^{\prime}\rightarrow \gamma\,U$ as an additional contribution mode for study of unparticle physics. }


\end {abstract}
PACS numbers: 13.25.Jx, 14.70.Pw




\bigskip

{\it Introduction.-} In 1982, Banks and Zaks [1]  investigated gauge theories containing non-integer
number of Dirac fermions where  the two-loop  $\beta$-function disappeares.
There is no chance to interpret theory at the non-trivial infra-red (IR) fixed point where it possesses the scale-invariant nature in terms of particles with definite masses. The main idea is based on the following statement: at very high
energies the theory contains both fields of the Standard Model (SM) and fields yielding
the sector with the IR point. Both of these sectors interact with each other by means
of exchange with the particles (fields) having a large mass $M$.
The hidden conformal sector may flow to IR fixed point at some scale $\Lambda < M$, where
the interaction between fields has the form
$\sim \Lambda ^{d_{BZ} - d}\,O_{SM}\,O_{U}\, M^{-a}$, where $a >0$,  $d_{BZ}$ and $d$ mean the scale dimensions
of the Banks-Zaks (BZ) sector operator and the operator $O_{U}$ of the $U$ - unparticle, respectively;
$O_{SM}$ is the operator of the SM fields.

 The unparticles (or the scale-invariant stuff) with continuous mass distribution, introduced
by H. Georgi in 2007 [2], obey the conformal (or scale) invariance.
There is an extensive literature (see, e.g., the incomplete set of papers [3] and
 the references therein) concerning the phenomenology of unparticles.

The most interesting scale is  $\Lambda\sim$ O(TeV), at which the
dynamics of unparticles could be seen at CERN LHC through the different
processes including the decays with the production of $U$- unparticles.

It has been emphasized [2] that the renormalizable interactions between the SM fields
and the fields of yet hidden conformal sector could be realized by means of explore the hidden energy
at high energy collisions and/or associated with the registration of non-integer number of
invisible particles. In this case, the conformal sector described in terms of  "unparticles" does not
possess those quantum numbers which are known in the SM.

Unparticle production at hadron colliders will be a signal that the scale where conformal invariance becomes important
for particle physics is as low as a few TeV. At this scale, the unparticle stuff sector is strongly coupled.
This requires that, somehow, a series of new reactions that involve unparticle stuff in an essential way
turn on between the Tevatron and LHC  energies. It will be important to understand this transition as precisely
as possible. This can be done through the study of $ p\bar{p}, pp \rightarrow \gamma + U$ and the identification
of the effects from, e.g., $Z-, Z^{\prime}-$ resonances [4] in $p\bar{p}, pp \rightarrow fermion + antifermion$.

The experimental channels of multi-gauge boson  production ensure the unique possibility to investigate
the anomalous triple effects of interaction between the bosons. We point out the study of
non-abelian gauge structure of the SM, and, in addition, the search for new types of interaction which,
as expected, can be evident at the energies above the electroweak scale. The triple couplings of neutral
gauge bosons, e.g., $ZZ^{\prime}\gamma$, $ZZZ^{\prime}$ etc. can be studied in pair
production at the hadron (lepton) colliders: $pp, p\bar{p} (e^{+}e^{-})\rightarrow
Z^{\prime}\rightarrow Z\gamma, ZZ,...$

In this paper, we study the production of unparticle $U$ in decays of $Z^{\prime}$ with a single photon emission.
There is a hidden sector where the main couplings to matter fields are
through the gauge fields. 
Before to use the concrete model, we have to make the following retreat.
First of all, we go to the extension of the Landau - Yang theorem [5,6] for the decay of a vector
particle into two vector states. Within this theorem, the decay of particle with spin-1 into two photons is
forbidden (because both outgoing particles are massless). The direct interaction between  $Z^{\prime}$ - boson
and a vector massive particle, e.g., $Z$ or $U$ - vector unparticle, accompanied by a photon, does not exist.
To the lowest order of the coupling constant $g$,  the contribution given by  $g^3$ in the decay
$Z^{\prime}\rightarrow \gamma U$ is provided mainly
by heavy quarks in the loop. What is the origin of this claim? First, it is worth to remember the
known calculation of the anomaly triangle diagram $ZZ\gamma$ [7], where the anomaly contribution result contains
two parts, one of which has no the dependence of the mass $m_{f}$ of (intermediate) charged fermions in the loop,
while the second part is proportional to  $m^{2}_{f}$. An anomaly term disappears in the case if all the fermions
from the same generation are taking into account or the masses of the fermions of each of generation
are equal to each other. The reason which explains the above mentioned note is the equality to zero of the sum
$\Sigma_{f} N^{f}_{c}\,g^{f}_{V}\,g^{f}_{A}\,Q_{f}$, where $g^{f}_{V} (g^{f}_{A})$ is the vector (axial-vector)
coupling constant of massive gauge bosons to fermions, $Q_{f}$ is the fermion charge,  $N^{f}_{c} = 3 (1)$
for quarks (leptons). The anomaly contribution for the decay $Z^{\prime}\rightarrow \gamma U$ does not
disappear due to  heavy quarks, and the amplitude of this decay is induced by the anomaly effect.
The contribution from light quarks with the mass  $m_{q}$ is suppressed as
 $m^{2}_{q}/m^{2}_{Z^{\prime}}\sim 10^{-8} - 10^{-6}$, where $m_{Z^{\prime}}$ is the mass of $Z^{\prime}$ - boson.
Despite the decay  $Z^{\prime}\rightarrow \gamma U$ is the rare process, there is a special attention
to the sensitivity of this decay to top-quark and even to quarks of fourth generation.

Since the photon has the only vector nature of interaction with the SM fields, the possible types
of interaction $Z^{\prime} - U - \gamma$ would be either $ V - A - V$ or  $ A - V - V$, where $V (A)$ means
the vector (axial-vector) interaction.

{\it Set up.-} Let us consider the following interaction Lagrangian density

\begin{equation}
\label{eq2}
-L = g_{Z^{\prime}}\sum_{q} \bar q (v^{\prime}_{q}\,\gamma^{\mu} -
a^{\prime}_{q}\,\gamma^{\mu}\,\gamma_{5}) q \,Z^{\prime}_{\mu} +
\frac{1}{\Lambda^{d-1}}\sum_{q} \bar q (c_{v}\,\gamma^{\mu} -  a_{v}\,\gamma^{\mu}\gamma_{5})
q\, O_{{\mu}_{U}},
\end{equation}
where $g_{Z^{\prime}} = (\sqrt {5b/3}\,s_{W}\,g_{Z})$ is the gauge constant of $U^{\prime}(1)$ group
(the coupling constant of $Z^{\prime}$ with a quark $q$) with the group factor $\sqrt {5/3}$,
$b\sim O(1)$, $g_{Z}=g/c_{W}$; $s_{W}(c_{W})= \sin\theta_{W} (\cos\theta_{W})$, $\theta_{W}$ is the angle of weak
interactions (often called as Weinberg angle); $v^{\prime}_{q}$ and $a^{\prime}_{q}$ are generalized vector and
the axial-vector  $U^{\prime}(1)$ -charges, respectively. These latter charges are dependent on both (joint)
gauge group  and the Higgs representation which is responsible for the breaking of initial gauge group to
the SM one; $c_{v}$ and $a_{v}$ are unknown vector and axial-vector couplings.
Actually, the second term in (\ref{eq2}) is identical to the first one up to the factor
$\Lambda^{1-d}$.

In conformal theory the unparticle does not have a fixed invariant mass, but instead has a continuous mass spectrum. 
In the paper we assume that  $O{_{\mu}}_{U}$ is a non-primary operator derived by $O{_{\mu}}_{U}(x) = \partial_{\mu} S(x)$ through the light  pseudo-Goldstone field $S(x)$ 
which is the consequence of an approximate continuous symmetry.  The  scalar field $S(x)$ is a "grandfather" potential which serves  as an approximate conformal compensator with continuous mass. The scale dimension of the gauge invariant non-primary vector operator is $d\geq 2$ as opposed to  $d\geq 3$ for primary gauge-invariant vector operators. In conformal theory, a primary operator defines the highest weight of representation of the conformal symmetry and this operator obeys the unitarity condition $d\geq j_{1} + j_{2} + 2 -\delta_{j_{1}j_{2}, 0}$, where $j_{1}$ and $j_{2}$ are the operator Lorentz spins (primary means not a derivative of another operator). For the review of the constraints of gauge invariant primary operators in conformal theory and the violations of the unitarity see [8] and the references therein. Furthermore, we consider $d\geq 1$ which does not contradict the unitarity condition because the operator $O{_{\mu}}_{U}(x) = \partial_{\mu} S(x)$  is not gauge invariant. Because the conformal sector is strongly coupled, the  mode $S(x)$ may be one of new states accessible at high energies.
 The operator $O{_{\mu}}_{U}$ has both the vector and the axial-vector couplings to quarks in the loop.


We consider the model containing the $Z_{\chi}$- boson on the scale
$O(1~TeV)$ in the frame of the symmetry based on the $E_{6}$ effective gauge group [9,10].
The coupling constant of $U(1)_{\chi}$ has the form $g_{\chi} = \sqrt{5/3}\,e/c_{W}$.

The amplitude of the decay $Z^{\prime}\rightarrow \gamma U$, where the coupling $Z^{\prime}\, U\,\gamma$
is supposed to be extended by the intermediate quark loop, has the form:
\begin{equation}
\label{eq4}
Am(z_{u},z_{q}) = \frac{e^{2}}{c_{W}}\sqrt{\frac{5}{3}}\frac{3}{\Lambda^{d-1}}
\sum_{q} e_{q}  \left ( c_{v}\,a^{\prime}_{q} +
 a_{v}\,v^{\prime}_{q}\right ) I(z_{u},z_{q})
\end{equation}
with $z_{u}=P^{2}_{U}/m^{2}_{Z^{\prime}}$,
$z_{q}=m^{2}_{q}/m^{2}_{Z^{\prime}}$ for the momentum $P_{U}$ of $U$ - unparticle
and the quarks  $q$ (in the loop) with the mass $m_{q}$.
We deal with the following expression for $I(z_{u},z_{q})$:
\begin{equation}
\label{eq44}
I=\frac{1}{1-z_{u}}\left \{\frac{1}{2}+\frac{z_{q}}{1-z_{u}}\left [F(z_{q}) -
F\left (\frac{z_{q}}{z_{u}}\right)\right]-\frac{1}{2(1-z_{u})}\left [G(z_{q}) -
G\left (\frac{z_{q}}{z_{u}}\right)\right]\right\}
\end{equation}
adopted for the decay  $Z^{\prime}\rightarrow \gamma U$ taking into account the results obtained in
[10] and [11].
For heavy quarks, $m_{q} > 0.5 \,m_{Z^{\prime}}$, the functions $F(x)$ and $G(x)$  in (\ref{eq44}) are
$$F(x)= - 2 {\left (\sin^{-1}\sqrt {\frac{1}{4x}}\right
)}^{2},\,\, G(x) = 2\sqrt {4x-1}\,\sin^{-1} \left (\sqrt
{\frac{1}{4x}}\right ), $$
while for light quarks ($m_{q} < 0.5 \,m_{Z^{\prime}}$), one has to use the formulas:
$$ F(x)= \frac{1}{2}{\left (\ln\frac{y^{+}}{y^{-}}\right)}^{2} + i\pi\ln\frac{y^{+}}{y^{-}}
-\frac{\pi^{2}}{2},\,\,\,\,
G(x) = \sqrt{1-4x} \left (\ln{\frac{y^{+}}{y^{-}}} + i\pi\right ),$$
where $y^{\pm} = 1 \pm\sqrt {1 -4\,x}$.
The variable $z_{u}$ is related to the photon energy $E_{\gamma}$  as $z_{u} = 1 - 2\,E_{\gamma}/m_{Z^{\prime}}$.
In the frame of the  $Z_{\chi}$ - model, we choose $v^{\prime}_{up} = 0,\, a^{\prime}_{up} = \sqrt{6}\,s_{W}/3,\, v^{\prime}_{down} =
2\,\sqrt{6}\,s_{W}/3,\, a^{\prime}_{down} = -\sqrt{6}\,s_{W}/3$ for $up$ - and $down$ - quarks. Actually,
 the account of the only light quarks leads to the zeroth result for the amplitude (\ref{eq4}).
In the heavy quark sector, the contribution $\sim a_{v}\,v^{\prime}_{q}$ is nonzero for the only $b$ - quarks and down - quarks of fourth generation. We emphasize that the nonvanishing result for the amplitude $Z^{\prime}\rightarrow \gamma U$ is the reflection of the anomaly contribution due to the presence of heavy quarks.



{\it Decay rate.-} In the decay $Z^{\prime}\rightarrow \gamma U $, the unparticle can not be identified with the definite invariant mass.
$U$- stuff possesses by continuous mass spectrum, and can not be in the rest frame (there is the similarity
to the massless particles). Since the unparticles are stable (and do not decay), the experimental signal of their
identification could be looking through the hidden (missing) energy and/or the measurement of the momentum distributions
when the $U$- unparticle is produced in $Z^{\prime}\rightarrow \gamma\,U$ decay.

The differential distribution of the decay width ��- $\Gamma(Z^{\prime}\rightarrow \gamma U)$ over the variable
$z_{u}$ looks like (see also [4]):
\begin{equation}
\label{eq5}
\frac{d\Gamma}{d z_{u}} = \frac{1}{2m_{Z^{\prime}}}\, \sum_{q} {\vert M_{q}\vert}^{2}\frac{A_{d}}
{16\,\pi^{2}}\left (m^{2}_{Z^{\prime}}\right )^{d-1}\, z^{d-2}_{u} (1-z_{u}),
\end{equation}
where
\begin{equation}
\label{eq6}
 \sum_{q} {\vert M_{q}\vert}^{2} = \frac{1}{6\,\pi^{4}}\, z_{u}\,(1-z_{u})^{2}\,(1+z_{u})\,
 {\vert Am(z_{u},z_{q})\vert}^{2}\, m^{2}_{Z^{\prime}},
\end{equation}
and [2]
\begin{equation}
\label{eq7}
A_{d} =\frac{16\,\pi^{5/2}}{(2\,\pi)^{2d}}\, \frac{\Gamma(d+1/2)}{\Gamma(d-1)\,\Gamma(2d)}.
 \end{equation}


One of the requirements applied to the amplitude in (\ref{eq6}) is that it
disappears in case of "massless" unparticle (Landau-Yang theorem), and when $z_{u} =1$.

Some bound regimes in (\ref{eq5}) may be both useful and instructive for further investigation.
For this, we consider the quark-loop couplings in the amplitude (\ref{eq4}) as the sum
of the contributions given by light quarks $q$ and heavy ones $Q$
($Q$ may be referred to the quarks of 4-th generation as well):
$$\sum_{q} e_{q}  \left (c_{v} a^{\prime}_{q} +
 a_{v}\,v^{\prime}_{q}\right ) I(z_{u},z_{q}) +
\sum_{Q} e_{Q}  \left ( c_{v} a^{\prime}_{Q} +
 a_{v}\,v^{\prime}_{Q}\right ) I(z_{u},z_{Q}).$$
For the SM quarks with $z_{q} < 1/4 $ and $(z_{q}/z_{u}) > 1/4 $ one gets the one-loop function $I(z_{u},z_{q})$
$$I(z_{u},z_{q})\simeq \frac{1}{2(1- z_{u})}\,\left (\frac{1}{3} -i\,\pi\right )\,(1+ z_{u}) $$
and the distribution ${d\Gamma}/{d z_{u}}$ has the form
$$\frac{d\Gamma}{d z_{u}} \simeq
\frac{5A_{d}}{6}\left (\frac{1}{9} + \pi^{2}\right )\left (\frac{s_{W}}{c_{W}}\right )^{2}
\left (\frac{\alpha}{2\,\pi^{2}}\right )^{2}\left (\frac{m^{2}_{Z^{\prime} }}{\Lambda ^{2}}\right )^{d-1}\,z_{u}^{d-1}
(1-z_{u}^{2})(1+ 2\,z_{u}).$$

On the other hand, if the quarks inside the loop become heavy enough, $z_{Q} > 1/4$, we estimate the
following  function $I(z_{u},z_{Q})$
$$I(z_{u},z_{Q}) \simeq \frac{1}{12\,z_{Q}\,(1-z_{u})}\left [\frac{1}{2}\,(1+ z_{u})\left (\frac{1}{8\,z_{Q}} -1\right ) +1 \right ], $$
which is very small in the limit $z_{Q} >> 1$.

The photon energy
$E_{\gamma} = m_{Z^{\prime}} (1- z_{u})/2$ in the limit $z_{u}\rightarrow 0$ gets its finite value.
The limit ${z_{u}\rightarrow 1}$ is trivial and we do not consider this.


Within the fact of the combination $ c_{v} a^{\prime}_{q} +  a_{v}\,v^{\prime}_{q}$
in (\ref{eq4}) the decay amplitude of $Z^{\prime}\rightarrow \gamma U$ does not disappear when the
summation on all the quarks degree of freedom is performed.

Constraints on the unparticle parameter $\Lambda$ can be obtained through, e.g., limits on measurable
collider phenomenology. In particular, it has been noted [12] that this bound never dips below 1 TeV.
In Figure 1, the monophoton energy distribution $E_{\gamma}^{-1}\,d\Gamma/dz_{u}$ is presented for a
range of photon energy  $E_{\gamma}$ = 0 - 500 GeV and various choices of $d$. For simplicity, we use the
$E_{\chi}$-model assuming the
flavor blind universality $c_{v} = a_{v} =1$ for all three generation quarks;
$\Lambda$ and $m_{Z^\prime}$ are set to be 1 TeV each.

\begin{figure}
  \centering
    \includegraphics[width=\textwidth, height = 75mm]{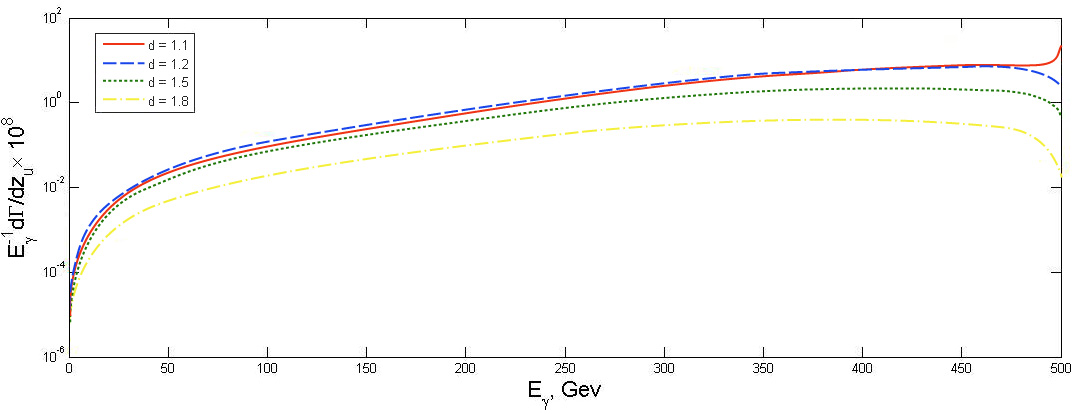}
    \caption{{ \it Energy distribution
    $E_{\gamma}^{-1}\,d\Gamma/dz_{u}\times 10^{8}$ for  $E_{\gamma} = 0 - 500$ GeV,
 depending on
$d= 1.1, 1.2, 1.5, 1.8$; $c_{v} = a_{v} =1$, $\Lambda$ = 1 TeV, $m_{Z^{\prime}} =$ 1 TeV.}}
  \label{fig:Ed}
\end{figure}


The sensitivity of the scale dimension $d$ to the energy distribution is evident. As $d$ moves away from unity, the energy spectrum
 begins to flatten out gradually, excepting $d = 1.2$ distribution which is above  $d = 1.1$. Such a behaviour is given by
 the factor (\ref{eq7}).

The $U$-unparticle could behave as a very broad vector boson since its mass could be distributed
over a large energy spectrum. The production cross-section into each energy bin could be much smaller
than in the case where a SM vector boson has that particular mass. This may be the reason why we have not yet
seen the  $U$-unparticle trace in the experiment.

In the appropriate approximation when the relation between the total decay width $\Gamma_{Z^{\prime}}$
of $Z^{\prime}$ - boson and
$ m_{Z^{\prime}}$ is small, the contribution to the cross section of the process
$pp\rightarrow Z^{\prime}\rightarrow \gamma\, U$ can be separated into $Z^{\prime}$ production cross section
$\sigma (pp\rightarrow Z^{\prime})$ and the branching ratio of the decay $Z^{\prime}\rightarrow \gamma\, U$,
$B(Z^{\prime}\rightarrow \gamma\, U) = \Gamma (Z^{\prime}\rightarrow\gamma\,U)/
\Gamma_{Z^{\prime}}$:
$\sigma (pp\rightarrow Z^{\prime}\rightarrow \gamma\, U) =  \sigma (pp\rightarrow Z^{\prime})\cdot
B(Z^{\prime}\rightarrow \gamma\, U)$.
The $Z^{\prime}$-boson can be directly produced at a hadron collider via the quark-antiquark annihilation
subprocess $\bar q q\rightarrow Z^{\prime}$, for which the cross section in the case of infinitely narrow
$Z^{\prime}$ is given by
\begin{equation}
\label{eq661}
\sigma (\bar {q}q\rightarrow Z^{\prime}) = k_{QCD}\,\frac{4\,\pi^{2}}{3}\,\frac{\Gamma (Z^{\prime}\rightarrow \bar {q}q)}
{m_{Z^{\prime}}}\,\delta (\hat s - m_{Z^{\prime}}^{2}),
\end{equation}
where $k_{QCD}\simeq$ 1.3 represents the enhancement from higher order QCD processes.
Conservation of the energy-momentum implies that the invariant mass of $Z^{\prime}$ is equal to the
parton center-of-mass energy $\sqrt{\hat {s}}$, with $\hat {s} = x_{1}x_{2}s$;
$x_{1}$ and $x_{2}$ are the fractions of the momenta carried by partons in the process
$\bar {q} q \rightarrow Z^{\prime}$.
The decay width $\Gamma (Z^{\prime}\rightarrow \bar {q}q)$ is
$$\Gamma (Z^{\prime}\rightarrow \bar {q}q) =  \frac{G_{F}\,m_{Z}^{2}}{6\pi\,\sqrt{2}}\,N_{c}\,m_{Z^{\prime}}\,
\sqrt {1- 4\,z_{q}}\,\left [\left (v^{\prime}_{q}\right )^{2} (1+ 2\,z_{q}) +  \left (a^{\prime}_{q}\right )^{2}(1- 4\,z_{q})\right ], $$
where $G_{F}$ is the Fermi coupling constant. In the narrow width approximation, the cross section (\ref{eq661}) reduces to
 ($z_{q} << 1$)
$$\sigma (\bar {q}q\rightarrow Z^{\prime}) \simeq k_{QCD}\,\frac{2\,a}{3}\,\frac{G_{F}}{\sqrt{2}}
\left (\frac{m_{Z}}{m_{Z^{\prime}}}\right )^{2}\frac{\left [\left (v^{\prime}_{q}\right )^{2} +
\left (a^{\prime}_{q}\right )^{2}\right ]}{(\hat {s}/m_{Z^{\prime}}^{2} -1 )^{2} + a^{2}},$$
where $a = \Gamma_ {Z^{\prime}}/m_{Z^{\prime}}$.

For the SM quarks, $ z_{q} \leq 1/4$, the branching ratio
$B (Z^{\prime}\rightarrow\gamma\,U)$ is
$$ B  = \frac{5A_{d}}{3a}\left (\frac{1}{9} + \pi^{2}\right )
 \left (\frac {s_{W}}{c_{W}}\right )^{2}
 \left (\frac{\alpha}{2\pi^{2}}\right )^{2}
  \left (\frac{m_{Z^{\prime}}^{2}}{\Lambda^{2}}\right )^{d-1}
  \left [\frac{1}{d(d+2)} +\frac{2}{(d+1)(d+3)}
  \right ]. $$
In Figure 2, the branching ratio $B (Z^{\prime}\rightarrow\gamma\,U)$  is presented with the assumption
$c_{v} = a_{v} =1$ for all three generation quarks;
$\Lambda$ is set to be 1 TeV, the range of  $d$ is chosen as = 1.1, 1.2, 1.5, 1.8 for
$m_{Z^{\prime}} =$ 0.5 -- 3.0 TeV. The width $\Gamma_{Z^{\prime}}$ is chosen
in the framework of the Sequential SM ($Z^{\prime}_{SSM}$) , where
the ratio $\Gamma_{Z^{\prime}}/m_{Z^{\prime}}$
has the maximal value $ a= 0.03$ among the Grand Unification Theories (GUT) inspired $Z^{\prime}$
models [13].

\begin{figure}
  \centering
  \includegraphics[width=\textwidth, height = 75mm]{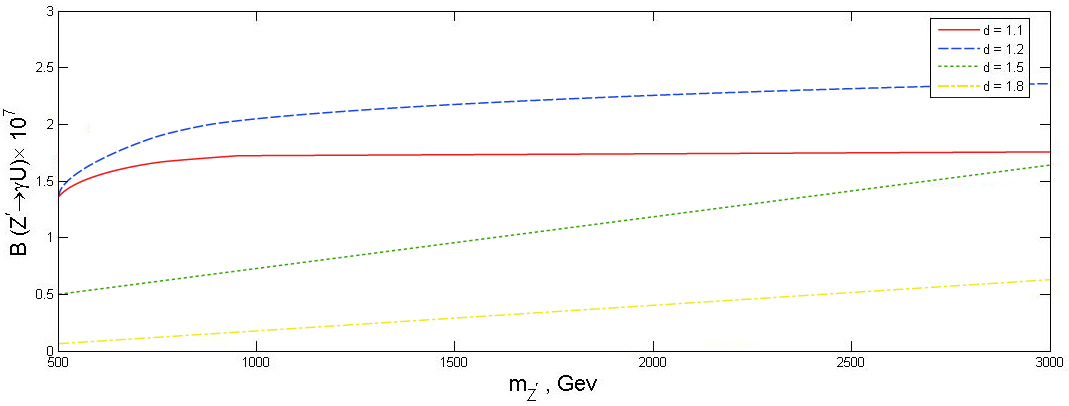}
    \caption{{ \it Branching ratio $B (Z^{\prime}\rightarrow\gamma\,U)\times 10^{7}$ for
$c_{v} = a_{v} =1$, $\Lambda =1$ TeV depending on
$d= $1.1, 1.2, 1.5, 1.8 and $m_{Z^{\prime}} =$ 0.5 -- 3.0 TeV with
$\Gamma_{Z^{\prime}}$ given by the $Z^{\prime}_{SSM}$-model ($a = 0.03$)}}
  \label{fig:animals}
\end{figure}

We find the smooth increasing of $B (Z^{\prime}\rightarrow\gamma\,U)$ with
$m_{Z^{\prime}}$ and its decreasing with the dimension $d$ excepting $d = 1.2$ branching ratio.\\
In Figure 3, we plot the cross section $\sigma (\bar q q\rightarrow Z^{\prime}\rightarrow \gamma\, U)$
in the case of up-quarks annihilation, where $x_{1}\sim x_{2}\sim \sqrt {x_{min}}$,
$x_{min} = m_{Z^{\prime}}^{2}/s$; the range of  $d$ is chosen as = 1.1, 1.2, 1.5, 1.8 for
$m_{Z^{\prime}} =$ 0.5 -- 3.0 TeV;  $ a= 0.03$.

\begin{figure}
  \centering
   \includegraphics[width=\textwidth, height = 75mm]{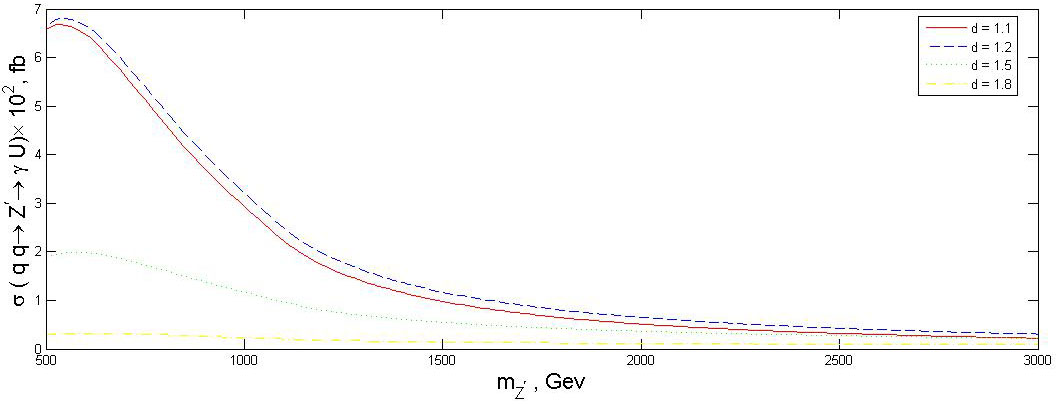}
    \caption{{ \it Cross section $\sigma (\bar q q\rightarrow Z^{\prime}\rightarrow \gamma\, U)\times 10^2, fb$
with the assumption of up-quarks annihilation, where $d$ = 1.1, 1.2, 1.5, 1.8,
$m_{Z^{\prime}} = $0.5 -- 3.0 TeV,  $ a= 0.03$.}}
  \label{fig:animals}
\end{figure}

For $100~fb^{-1}$ luminosity at the LHC, the detection of the process  $Z^{\prime}\rightarrow \gamma\, U$ can be achieved with
about 10 signal events for $m_{Z^{\prime}}\sim O(1~TeV)$ at $d = 1.1$.

{\it Constraints.-} The hidden sector can be strongly constrained by existing experimental
data. One of the important and practical implications for unparticle phenomenon is the analysis of the operator form
\begin{eqnarray}
\label{eq10}
  \frac{\Lambda^{d_{BZ} - d}}{M^{d_{BZ} - 2}}\,\vert H\vert^{2}\,O_{U}
\end{eqnarray}
containing the SM Higgs field $H$.
Within the Higgs vacuum expectation value ($v$)
requirement, the theory becomes non-conformal below the scale
$$\tilde\Lambda = \left (\frac{\Lambda^{d_{BZ} - d}}{M^{d_{BZ} - 2}}\,v^{2}\right )^
{\frac{1}{4-d}} < \Lambda, $$
where $U$-unparticle sector becomes a standard sector. For practical
consistency we require $\tilde\Lambda < \sqrt {s}$. It implies that unparticle physics phenomena
can be seen at high energy experiment with the energies
$$s > \left (\frac{\Lambda^{d_{BZ} - d}}{M^{d_{BZ} - 2}}\,v^{2}\right )^
{\frac{2}{4-d}}$$
even when $d\rightarrow d_{BZ}$. Note, that any observable involving operators
$O_{SM}$ and $O_{U}$ in (\ref{eq10}) will be given by the operator
$$\hat o = \left (\frac{\Lambda^{d_{BZ} - d}}{M^{d_{BZ} + n - 4}}\right )^
{2}\, s^{ d + n -4}, $$
where $ n$ is the dimension of the SM operator. Then, the observation of the unparticle
sector is bounded by the minimal energy
\begin{eqnarray}
\label{eq11}
s > {\hat o}^{\frac{1}{n}}\, M^{2}\, \left (\frac{v}{M}\right )^{\frac{4}{n}}.
\end{eqnarray}
No both $d$- and $d_{BZ}$ - dimensions we have in the lower bound (\ref{eq11}). The main
model parameter is the mass $M$ of heavy messenger. If the experimental deviation from
the SM is detected at the level of the order $\hat o \sim 1 \%$ at $n = 4$, the
lower bound on $\sqrt {s}$ would be from 0.9 TeV to 2.8 TeV for $M's$ running from
10 TeV to 100 TeV, respectively. Thus, both the Tevatron and the LHC are the ideal
colliders where the unparticle physics can be tested well.

{\it Conclusion.-}  We  studied the decay of the extra neutral gauge boson $Z^{\prime}$ into a vector $U$-unparticle and a photon. Both vector and axial-vector couplings to quarks play a significant role. 
The energy distribution for $pp\rightarrow Z^{\prime}\rightarrow \gamma\, U$ can discriminate $d$. The branching ratio $B (Z^{\prime}\rightarrow\gamma\,U)$ is at best of the order of $10^{-7}$ for the scale dimension $d = 1.1$. For larger $d$, the branching ratio is at least smaller by one order of the magnitude. Unless the LHC can collect a very large sample of $Z^{\prime}$-bosons, detection of $U$ through the decay  $Z^{\prime}\rightarrow\gamma\,U$ would be challenging compared to the decay $Z\rightarrow\gamma\,U$, where the  branching ratio $B (Z\rightarrow\gamma\,U) \sim10^{-8}$ [4].  

For  $100~fb^{-1}$ integrated luminosity the detection of  $Z^{\prime}\rightarrow \gamma\,U$ can be with about 10 signal events at $d = 1.1$ for a 1.0 TeV $Z^{\prime}$, while for larger values of $d$ there is the decreasing of the events number.



For the case when $Z^{\prime}$ - boson has continuously distributed mass [14], the branching ratio has an additional suppression factor due to nonzero internal deceay width  $\Gamma_{Z^{\prime}}^{int}$ in formulas: 
$$\int_{0}^{\infty} \frac{\rho (t)\,dt}{p^{2} - t- i\,\epsilon} \simeq \frac{1}
{p^{2} - m_{Z^{\prime}}^{2} + i\,m_{Z^{\prime}}\, \Gamma_{Z^{\prime}}^{int}}, \,\,\,
 \rho (t) =\frac{1}{\pi}\frac{\Gamma_{Z^{\prime}}^{int}\, m_{Z^{\prime}}}{(t-m^{2}_{Z^{\prime}})^{2} +  \Gamma^{2\,int}_{Z^{\prime}}\, m^{2}_{Z^{\prime}}} .$$
The experimental estimation of $B (Z^{\prime}\rightarrow \gamma\,U)$ could provide with the quantity $\Gamma_{Z^{\prime}}^{int}$, and since the $\gamma$-quantum energy has a continuous spectrum, by measuring the photon energy spectrum in the $Z^{\prime}$- decay, one can discriminate the presence of the $U$-unparticle or not.

We have shown numerical results for $Z^{\prime}$ -bosons associated with the $Z_{\chi}$ - model. The calculations are easily applicable to other extended gauge models, e.g., Little Higgs scenario models, Left-Right Symmetry Model, Sequential SM.






\end{document}